\documentclass[pdflatex,sn-mathphys-num]{sn-jnl}

\usepackage{graphicx}%
\usepackage{multirow}%
\usepackage{amsmath,amssymb,amsfonts}%
\usepackage{amsthm}%
\usepackage{mathrsfs}%
\usepackage[title]{appendix}%
\usepackage{xcolor}%
\usepackage{textcomp}%
\usepackage{manyfoot}%
\usepackage{booktabs}%
\usepackage{algorithm}%
\usepackage{algorithmicx}%
\usepackage{algpseudocode}%
\usepackage{listings}%

\theoremstyle{thmstyleone}%
\theoremstyle{thmstyletwo}%

\theoremstyle{thmstylethree}%

\begin{document}

\title[]{Multi-strange and charmed hadrons: A novel probe for the QCD equation of state at high baryon densities}


\author*[1,2]{\fnm{Jan} \sur{Steinheimer}}\email{j.steinheimer-froschauer@gsi.de}

\author[2,3,4]{\fnm{Tom} \sur{Reichert}}

\author[3,1,4]{\fnm{Marcus} \sur{Bleicher}}

\affil[1]{\orgname{GSI Helmholtzzentrum f\"ur Schwerionenforschung GmbH}, \orgaddress{\street{Planckstr. 1}, \city{Darmstadt}, \postcode{D-64291}, \country{Germany}}}

\affil[2]{\orgname{Frankfurt Institute for Advanced Studies}, \orgaddress{\street{Ruth-Moufang-Str. 1}, \city{Frankfurt am Main}, \postcode{60438}, \country{Germany}}}

\affil[3]{\orgdiv{Institut für Theoretische Physik}, \orgname{Goethe-Universit\"{a}t Frankfurt}, \orgaddress{\street{Max-von-Laue-Str. 1}, \city{Frankfurt am Main}, \postcode{60438}, \country{Germany}}}

\affil[4]{\orgname{Helmholtz Research Academy Hesse for FAIR (HFHF), GSI Helmholtzzentrum f\"ur Schwerionenforschung GmbH}, \orgaddress{\street{Max-von-Laue-Str. 12}, \city{Frankfurt am Main}, \postcode{60438}, \country{Germany}}}

\abstract{Nuclear experiments near and below the threshold of hyperon production have shown that the production of Kaons is a sensitive probe for the dense QCD equation of state. At beam energies up to 1.5AGeV, strangeness production can probe the equation of state for densities up to approximately twice nuclear saturation. In this paper we will discuss the possibilities of extending this range in density by the study of multi-strange baryons as well as charmed hadrons in the SIS100 beam energy range up to $10A$GeV. Here, densities up to five times nuclear saturation can be reached and the production of multi-strange and charmed hadrons shows a strong sensitivity to the equation of state. On the other hand a precise prediction of the effect of the equation of state will require knowledge of the fundamental production cross section near the elementary production threshold in p+p collisions which is yet not measured for the hadrons discussed.}




\maketitle

\section{Introduction}\label{sec1}

The production of strange hadrons has early been suggested as probe for the properties of the hot and dense QCD matter produced in relativistic heavy ion collisions \cite{Koch:1986ud,Gazdzicki:1996pk,Bass:1998vz,Becattini:2003wp}. While at the highest beam energies strange quarks are produced as pairs in a deconfined state, at lower energies the associated production via the excitation and decay of baryonic resonances ($N^* \rightarrow Y+K$) dominates the production. The total amount of ${s+\overline{s}}$ pairs produced in a nuclear reaction also depends strongly on the incident beam energy. While in central high energetic heavy ion collisions, the system allows for copious production of strangeness, its production becomes suppressed for smaller systems and beam energies below $\sqrt{s_{\mathrm{NN}}}\leq 5$ GeV. In a transport model this is understood as a result of the secondary reactions which can occur if a system of large enough size and energy is created, while in a statistical model picture this effect is usually attributed to the canonical effect \cite{Cleymans:1990mn}. 

Another interesting aspect of large collision systems is that it allows the production of strange (and charmed) hadrons below their elementary threshold, i.e. below their threshold energy in p+p collision systems. Again, this is only possible due to secondary interactions in the fireball produced by the heavy ion collisions.

Past studies, at beam energies below $E_{\mathrm{lab}}< 2A$ GeV, have shown that the sub-threshold production of kaons and hyperons is sensitive to the compression reached during the collisions and therefore sensitive to the equation of state of dense QCD matter \cite{Hartnack:1993bq,Hartnack:1993bp,Fuchs:2005zg,Hartnack:2005tr,Hartnack:2011cn}.

The beam energies and luminosities achieved at the upcoming SIS100 accelerator and CBM experiment are well suited to open up production thresholds for several (multi-)strange hadrons as well as charmed hadrons \cite{Steinheimer:2016jjk,Reichert:2025iwz}, for previous charm studies with UrQMD we refer to \cite{Spieles:1999kp,Lang:2012nqy}. Table \ref{tab:thresh} shows a list of these hadrons together with their elementary threshold energy. 

The idea of this work is to show the sensitivity of the (sub-threshold) production yield of strange and (for the first time) charm hadrons on the equation of state and discuss other uncertainties which will become important in the interpretation of the future measured multiplicities.

\begin{table}[h!]
\centering
\begin{tabular}{|c|c|}
\hline
Hadron   & Threshold $\sqrt{s_{\mathrm{NN}}}$ in p+p [GeV] \\
\hline
$\Lambda$ and Kaon & 2.55\\
\hline
anti-Kaon & 2.87\\
\hline
 $\Xi$ & 3.25\\
\hline
$\Omega$ & 4.10\\
\hline
$\Lambda_C$ & 5.09\\
\hline
$J/\Psi$ & 4.97\\
\hline
\end{tabular}
\caption{Approximate threshold energies for the production of different strange and charmed hadrons in elementary p+p reactions. \label{tab:thresh}}
\end{table}

\section{The UrQMD model}

UrQMD is a microscopic transport model propagating hadrons in phase-space according to their relativistic momenta \cite{Bass:1998ca,Bleicher:1999xi,Bleicher:2022kcu}. The equations of motion of hadrons in UrQMD are influenced by a binary scattering term and long range QMD potentials. The binary scattering is simulated in the cascade part of the model in which hadrons propagate on straight trajectories (unless long range QMD potentials are included) until they undergo an elastic or inelastic scattering which will change their momenta. The probability of such a scattering is given by a geometric interpretation of the scattering cross section. These cross sections are the fundamental input of the model and are taken (where available) from experimental measurements (e.g. \cite{ParticleDataGroup:2020ssz}) or from theoretical calculations (see \cite{Bass:1998ca,Bleicher:1999xi,Bleicher:2022kcu} for more details). UrQMD includes a comprehensive list of hadronic resonances that can be excited by inelastic scattering and thus the cascade mode of the model will resemble a system with hadron resonance gas (HRG) equation of state. Such an equation of state can be considered much softer than the usual nuclear equation of state implemented in the QMD part. In fact is was shown \cite{Steinheimer:2022gqb}, that the cascade mode can be considered almost as soft as if a phase transition was present during the evolution.

\begin{figure} [t]
    \centering
    \includegraphics[width=0.7\columnwidth]{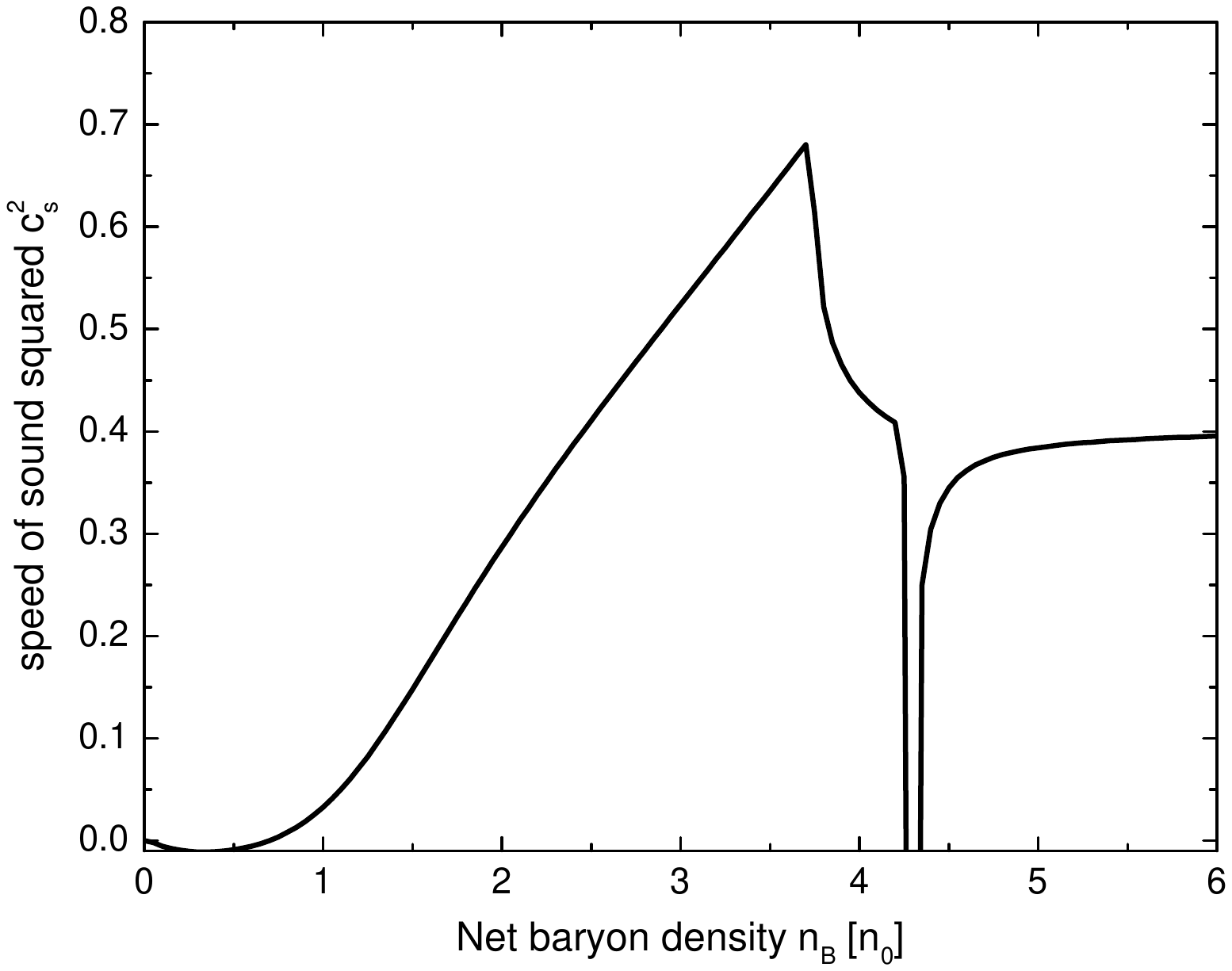}
    \caption{Speed of sound squared of the CMF parametrization that is used for the UrQMD simulations.}
    \label{fig:cs2_cmf}
\end{figure}

For a dense hadronic system, the HRG does not provide a complete description of the equation of state and this additional interactions can be included in the QMD-part of the model. From a recent development, UrQMD v4.0 now includes density and momentum dependent potentials for all baryons, based on a chiral mean-field (CMF) equation of state that is consistent with neutron star observations and lattice QCD results \cite{Motornenko:2019arp,Steinheimer:2024eha}. 
The CMF equation of state is based on an effective chiral Lagrangian and incorporated parity doubling for baryons allowing for a description of chiral symmetry restoration in dense matter. The baryons in the CMF interact via scalar and vector fields and thus allow a consistent calculation of the density and momentum dependent potentials used for the UrQMD simulations. The resulting speed of sound (at $T=0$) of the CMF version used is shown in figure \ref{fig:cs2_cmf}.
It was shown that with this equation of state the model is able to describe flow and hadron production at SIS18 energies \cite{Steinheimer:2025hsr}.

\subsection{Hadron production in UrQMD}\label{production}

The production of new hadrons, including strange and charmed states, in the UrQMD model can occur through three different channels:
\begin{enumerate}
\item The excitation and decay of a (baryonic) resonance. This is by far the dominant contribution to particle production in the SIS100 beam energy range. For example the binary scattering of a target and projectile nucleon will lead to the excitation of a baryonic resonance which then decays to strange hadrons. In addition, secondary scatterings can lead to the excitation of resonances through Meson+Baryon reactions. The cross section for the production of strange and charm hadrons then is a result of a time dependent folding of the cross section for the resonance excitation and corresponding branching ratio of the resonance decay. Both can be taken from experimental data for elementary reactions where available. The relevant cross sections for charm hadrons have been fixed as discussed in \cite{Steinheimer:2016jjk}
\item The excitation and decay of a color string. As the beam energy increases, the incoming nucleons will interact by forming a color field, leading to the observed energy loss. This color string will eventually fragment into quark+anti-quark (or di-quark) pairs which finally form new hadrons. The properties of the produced quarks and consequently hadrons is determined by the string fragmentation parameters which also includes a suppression for the production of strange quarks (motivated by the mass difference between u/d- and s-quarks). Again, the parameters for the string excitation and decay are determined by comparisons to elementary scattering data at higher beam energies. One should note that while the color string fragmentation in UrQMD produces strange quark pairs, charm production from the string is very strongly suppressed due to the very large mass difference between s- and c-quarks and currently not treated.
\item Flavor exchange reactions. After the original creation of strange hadrons they may further interact with other hadrons leading to the exchange of strange quarks between them. Such processes are for example $\Lambda + \pi \leftrightarrow N + \overline{K}$ which can occur either via the intermediate excitation of a baryonic resonance or the direct exchange reaction. While the first is naturally included via the resonance branching fractions and detailed balance, the latter was introduced in \cite{Graef:2014mra}. Here, one has to distinguish between meson-baryon and baryon-baryon (e.g. $\Lambda + \Lambda \leftrightarrow \Xi + N$) exchange reactions. The current version of UrQMD enables only meson-baryon exchange reactions while the baryon-baryon reactions are turned off by default due to lack of knowledge to benchmark the cross sections.
\end{enumerate}

\begin{figure} [t]
    \centering
    \includegraphics[width=0.7\columnwidth]{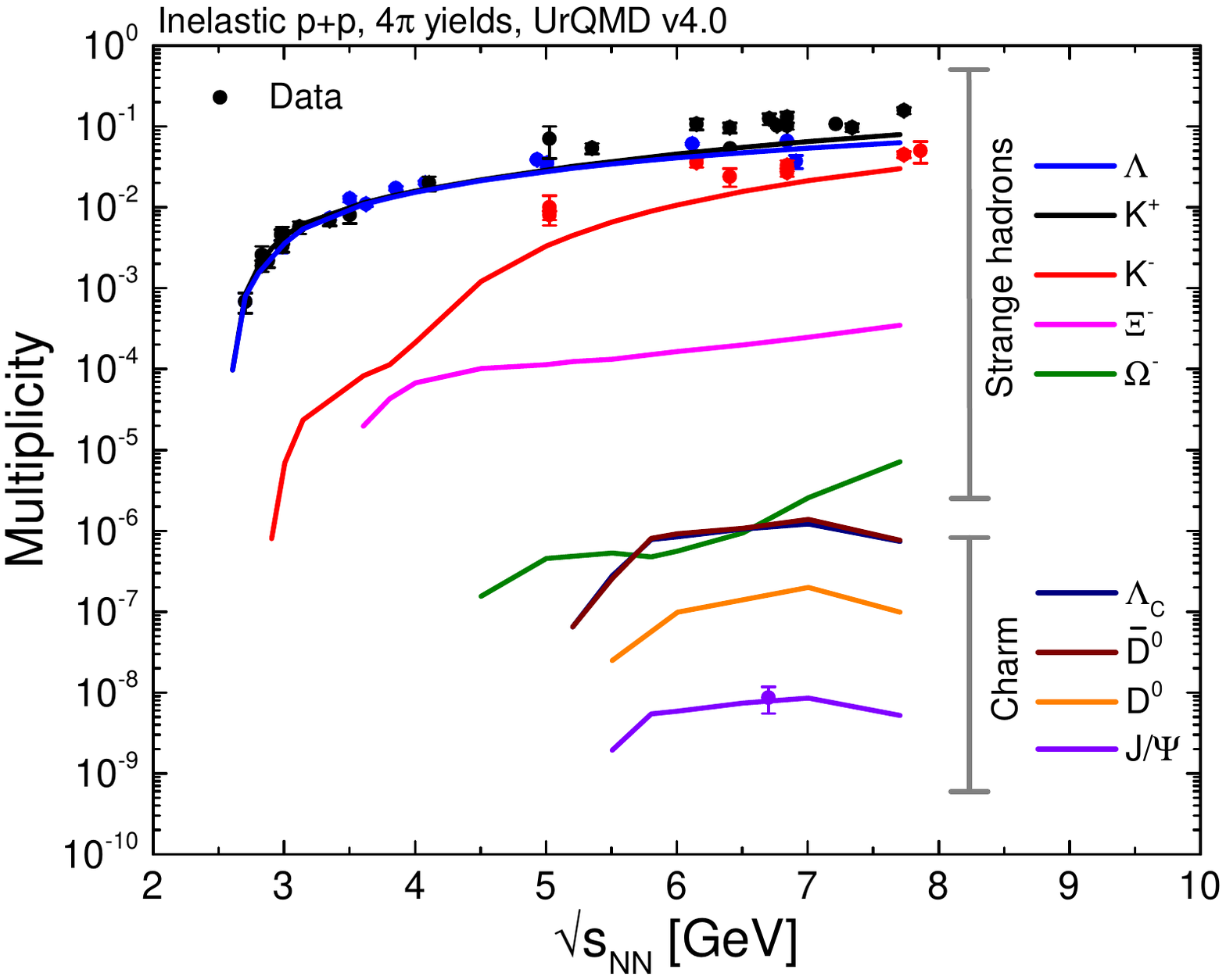}
    \includegraphics[width=0.7\columnwidth]{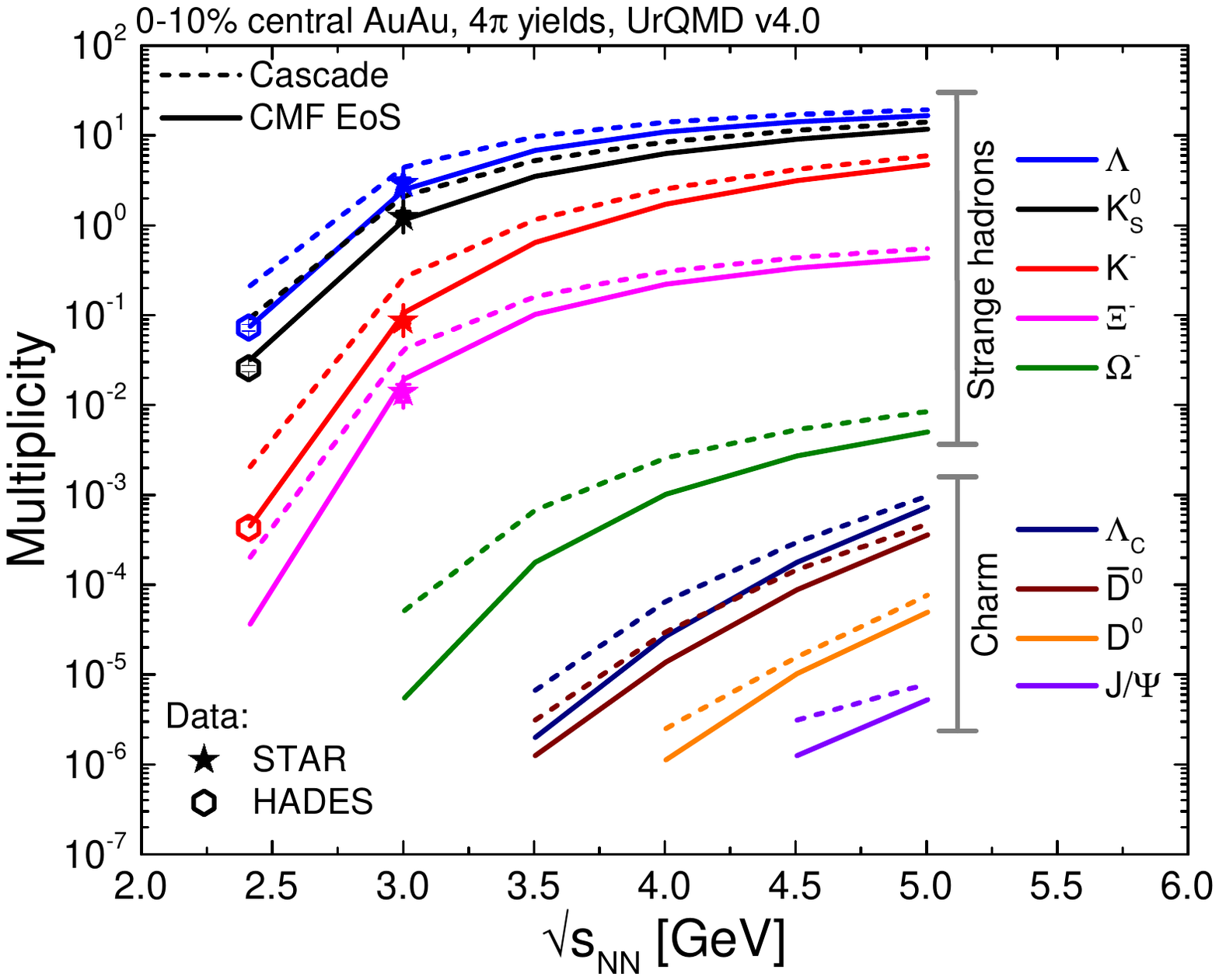}
    \caption{Upper panel: Integrated $4 \pi$ multiplicity per inelastic p+p event for several strange and charmed hadrons from UrQMD v4.0 (solid lines) and compared to experimental data \cite{Antinucci:1972ib,Rossi:1974if}. Lower panel: Multiplicities per event for central (0-10$\%$) Au+Au reactions with UrQMD (lines) and compared to experimental data from HADES and STAR (symbols) \cite{HADES:2017jgz,Li:2025yhe}. The solid lines correspond to simulations with the CMF-equation of state (CMF) while the dashed lines are results with the UrQMD cascade mode (CAS)}
    \label{fig:mul_pp}
\end{figure}

\section{Results}\label{sec2}

Figure \ref{fig:mul_pp} shows the production multiplicity of several strange and charmed hadrons from UrQMD v4.0 compared to available data. The upper panel of figure \ref{fig:mul_pp} shows the integrated multiplicity per inelastic p+p event as function of the center of mass energy. This is an important baseline for heavy ion reactions, as the the production cross sections in the transport model are often tuned to describe these elementary production rates. It is obvious, that there is a significant amount of experimental data for the production yields of kaons and Lambda hyperons (blue and black symbols) and fewer data for anti-kaons (red symbols). Where data is available, the transport model gives a reasonably good description. On the other hand, for the multi-strange hyperons and charmed hadrons there is essentially no data available in this beam energy range. 

\begin{table}[b]
\centering
\begin{tabular}{|c|c|c|}
\hline
Center of mass energy [GeV/c]   & Centrality & Number of participants $A_{\mathrm{part}}$ \\
\hline
2.41 & 0-10 $\%$ & 303 \\
2.41 & 10-20 $\%$ & 213 \\
2.41 & 20-30 $\%$ & 150 \\
2.41 & 30-40 $\%$ & 103 \\
\hline
3.0 - 5.0 & 0-10 $\%$ & 310 \\
3.0 - 5.0 & 10-20 $\%$ & 224 \\
3.0 - 5.0 & 20-30 $\%$ & 160 \\
3.0 - 5.0 & 30-40 $\%$ & 111 \\
\hline
\end{tabular}
\caption{Number of participants used for the different centralities. \label{tab:central}}
\end{table}

The lower panel of figure \ref{fig:mul_pp} shows the integrated multiplicities of different strange and charm hadrons in central ($0-10\%$) Au+Au collisions as function for the beam energy in the energy range of the SIS100 accelerator. The solid lines correspond to results with the CMF equation of state and the dashed lines are obtained from UrQMD in cascade mode (i.e. the QMD potentials are turned off). As expected there is a clear difference, depending on the equation of state used. Especially below their elementary threshold energy, particle yields are significantly suppressed in the CMF scenario due to the reduced maximal compression. The UrQMD simulations are also compared to available experimental data shown as colored symbols. Here, the simulations with the CMF-equation of state give a good description of the available data.

\begin{figure} [t]
    \centering
    \includegraphics[width=0.75\columnwidth]{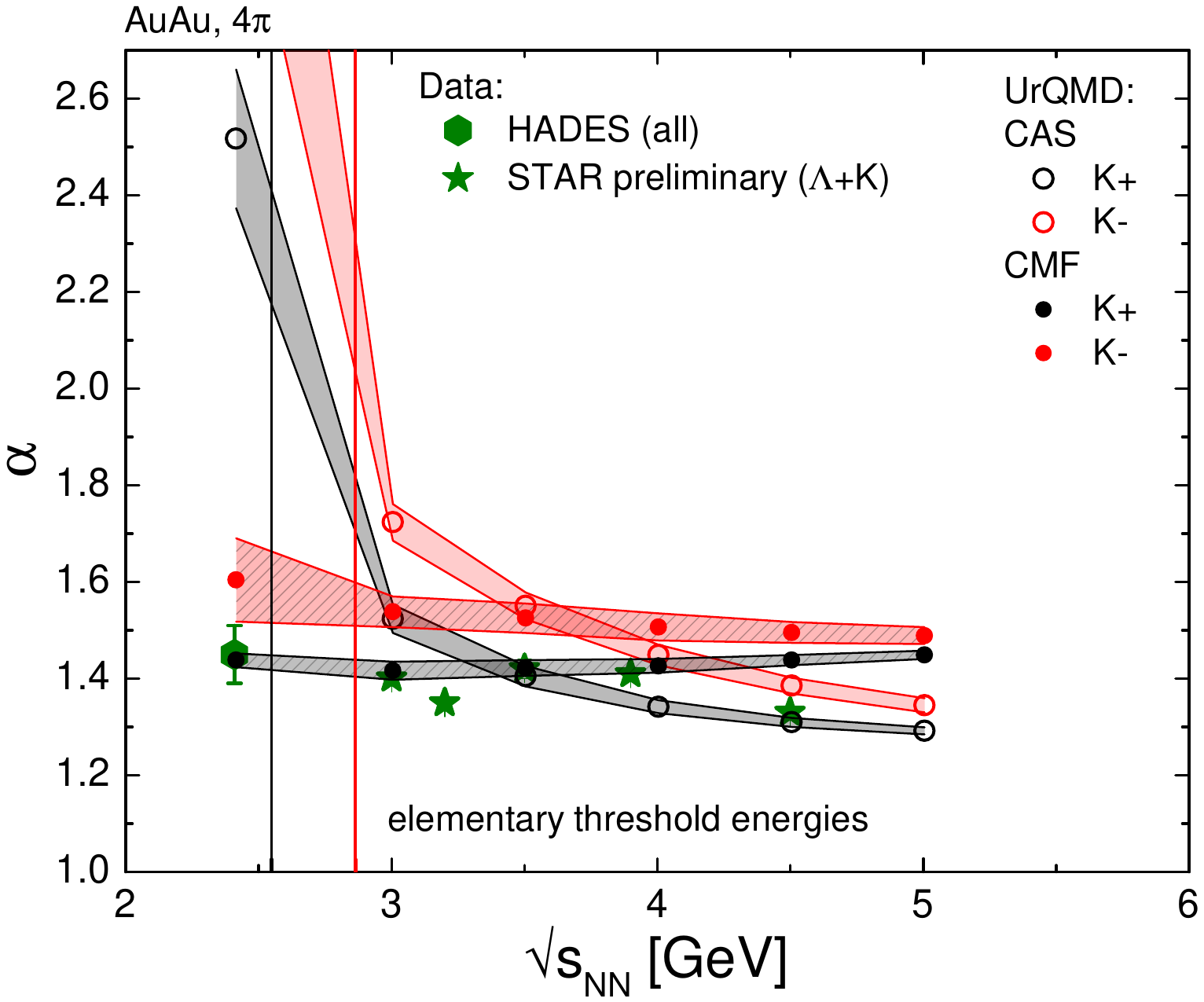}
    \caption{Beam energy dependence of the $\alpha$ parameter from the fit to the centrality dependence of Kaons and anti-Kaons in Au+Au collisions. The shaded areas with filled symbols denote the UrQMD results with the CMF equation of state while the open symbols are the results with the cascade mode (CAS). Experimental data for Kaons are shown as green symbols. The threshold energies for the production in elementary p+p collisions are indicated as vertical lines.}
    \label{fig:alpha_k}
\end{figure}

\begin{figure} [t]
    \centering
    \includegraphics[width=0.75\columnwidth]{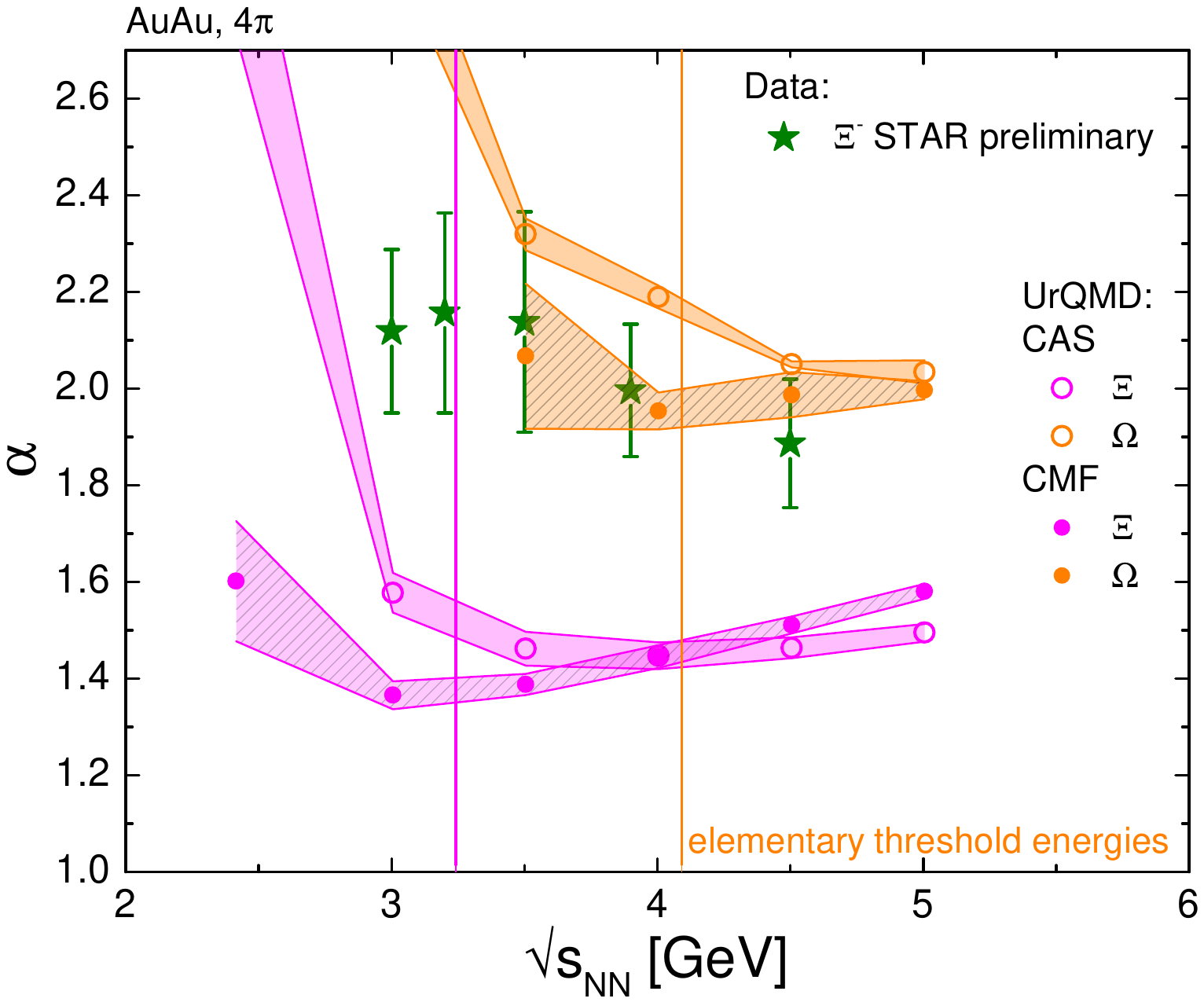}
    \caption{Beam energy dependence of the $\alpha$ parameter from the fit to the centrality dependence of $\Xi^-$ and $\Omega^-$ in Au+Au collisions. The shaded areas with filled symbols denote the results from UrQMD with the CMF equation of state while the open symbols are the results with the cascade mode (CAS). Preliminary experimental data for $\Xi^-$ are shown as green symbols. The threshold energies for the production in elementary p+p collisions are indicated as vertical lines.}
    \label{fig:alpha_xi}
\end{figure}

The strong dependence of the sub-threshold production of hadrons on the equation of state has been discussed in previous works \cite{Aichelin:1987ti,Hartnack:1993bq,Hartnack:2005tr,Hartnack:2011cn,Hong:2013yva}, where the focus was mainly on kaons and single strange hyperons. In particular the centrality dependence of strangeness production was shown to be a sensitive probe of the equation of state. The dependence of the hadron multiplicity $M^i$, scaled by the number of participant nucleons $A_{\mathrm{part}}$ in a given centrality bin, as function of the number of participants can be characterized by the parameter $\alpha_i$:

\begin{equation}
M^i(A_{\mathrm{part}}) = M^{i}_{0} \ A_{\mathrm{part}}^{\alpha_i}
\end{equation}

The parameter $\alpha$ quantifies the scaling of particle yield with the number of participants which can be related to the total system size but also to the degree of compression reached in a collision.
Previous studies showed that a larger value of $\alpha$ corresponds to a softer equation of state, as the larger compression in central collisions will lead to more secondary interactions, driving the strangeness yield towards equilibrium. In such a scenario, a significantly increased $\alpha$ signals the presence of production mechanisms in the high-density environment of central collisions.
On the other hand, the value of $\alpha$ may also depend critically on the specific in-medium production and absorption cross sections.

In the past, the parameter $\alpha$ was investigated only for single strange hadrons like Kaons and $\Lambda$ for which the threshold energy is low. Therefore, one was restricted to studies of the density dependence at lower beam energies.
Figure \ref{fig:alpha_k} shows the $\alpha$ for Kaons and anti-Kaons, extracted from UrQMD simulations with the two different equations of state, CMF (shades area with filled symbols) and cascade mode (denoted as CAS, open symbols). To compare the results of our simulations to experimental results, the centralities and number of participants where defined by the experiments Glauber fits to charged particles produced in Au+Au collisions in this energy range. The corresponding values are shown in table \ref{tab:central} and differ slightly between experiments. The parameter $\alpha$ was then extracted from a fit to the 4 most central centrality bins as shown in table \ref{tab:central}. The UrQMD values are also compared to results from the HADES and STAR collaborations. It is found that the CMF version of UrQMD gives a good description of the $\alpha$ parameter for Kaons over a wide energy range. As the cascade version of the model has a much softer equation of state, the $\alpha$ parameter increases drastically below the elementary threshold as expected. We also observe that for the $K^-$, since the threshold is higher than for the $K^+$, the increase is even stronger.

\begin{figure} [t]
    \centering
    \includegraphics[width=0.75\columnwidth]{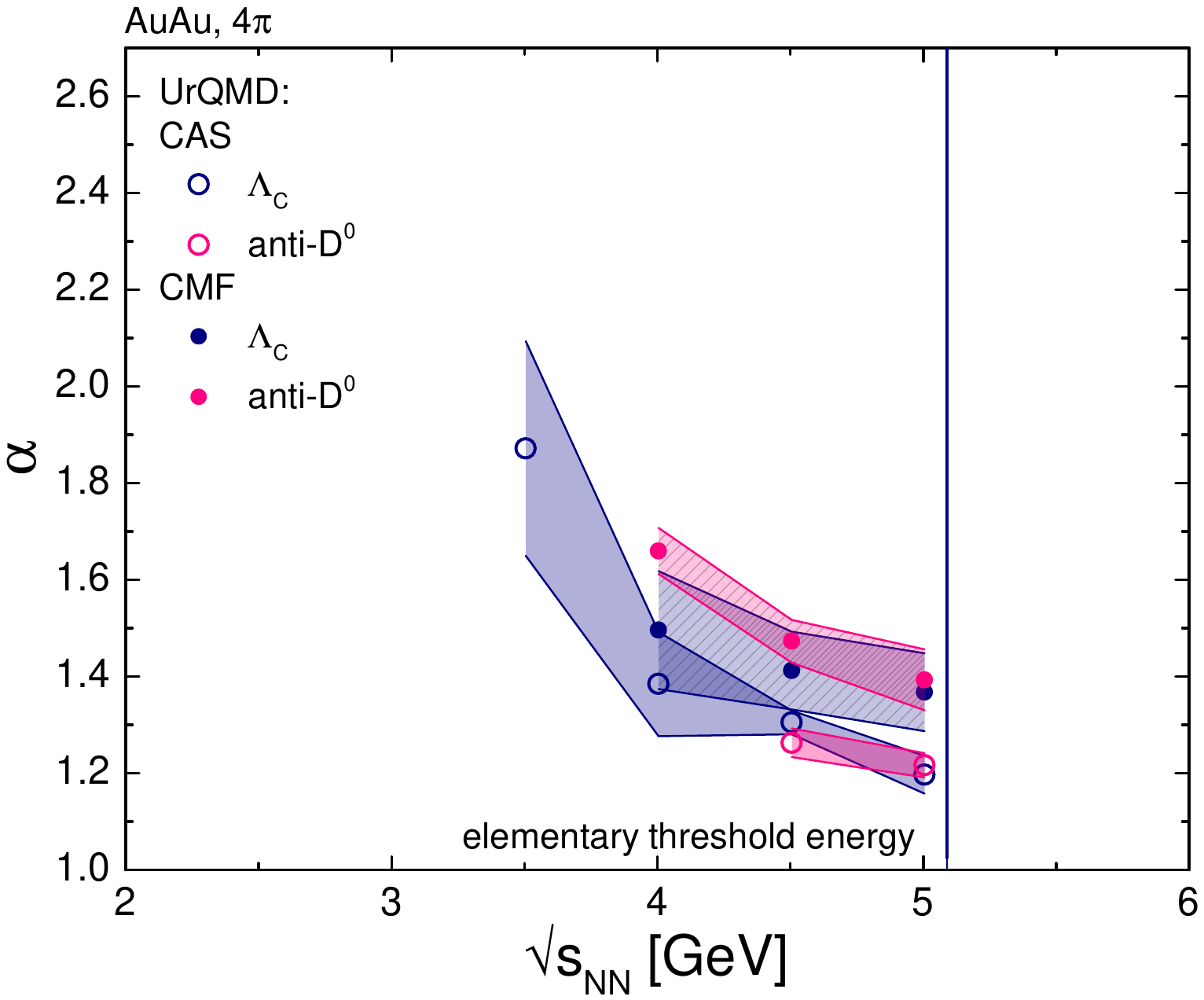}
    \caption{Beam energy dependence of the $\alpha$ parameter from the fit to the centrality dependence of charm hadrons in Au+Au collisions. The shades areas with filled symbols denote the results with the CMF equation of state while the open symbols are the results with the cascade mode (CAS). The threshold energy for production in elementary p+p collisions is indicated as vertical line.}
    \label{fig:alpha_c}
\end{figure}

The threshold energy for multi-strange hyperons and even charmed hadrons is much higher (given in table \ref{tab:thresh}) and therefore one can conjecture that the $\alpha$ parameter for such hadrons would be sensitive to the higher densities reached at higher beam energies. In figures \ref{fig:alpha_xi} and \ref{fig:alpha_c} we show the $\alpha$ parameter for multi-strange $\Xi$ and $\Omega$ baryons and for the charmed $\Lambda_C$ and $\overline{D^0}$ as function of beam energy. Again, results with the CMF equation of state are compared to cascade simulations (CAS). For the multi-strange hyperons the results are as expected - below the elementary threshold, the softer equation of state leads to a drastic increase of the $\alpha$ parameter. However, a comparison with preliminary STAR data for the $\Xi$, highlights an important complication. As we can see, the STAR data shows a value for $\alpha_{\Xi}$ which is significantly larger than observed in either UrQMD simulation, CAS and CMF. This emphasizes that $\alpha$ parameter may be influenced by further contributions and not just the equation of state.

Furthermore, the charm production shows a behavior which is essentially opposite to the strange hadrons. Here, the softer equation of state leads to a smaller $\alpha$ parameter even though the total production is increased. This is likely due to the fact that no additional charm producing processes are included in the simulation which would enhance charm production in dense systems, but the heavy resonances which eventually decay into charmed hadrons may be easier absorbed in the dense medium.

\subsection{Discussion}

\begin{figure} [t]
    \centering
    \includegraphics[width=0.75\columnwidth]{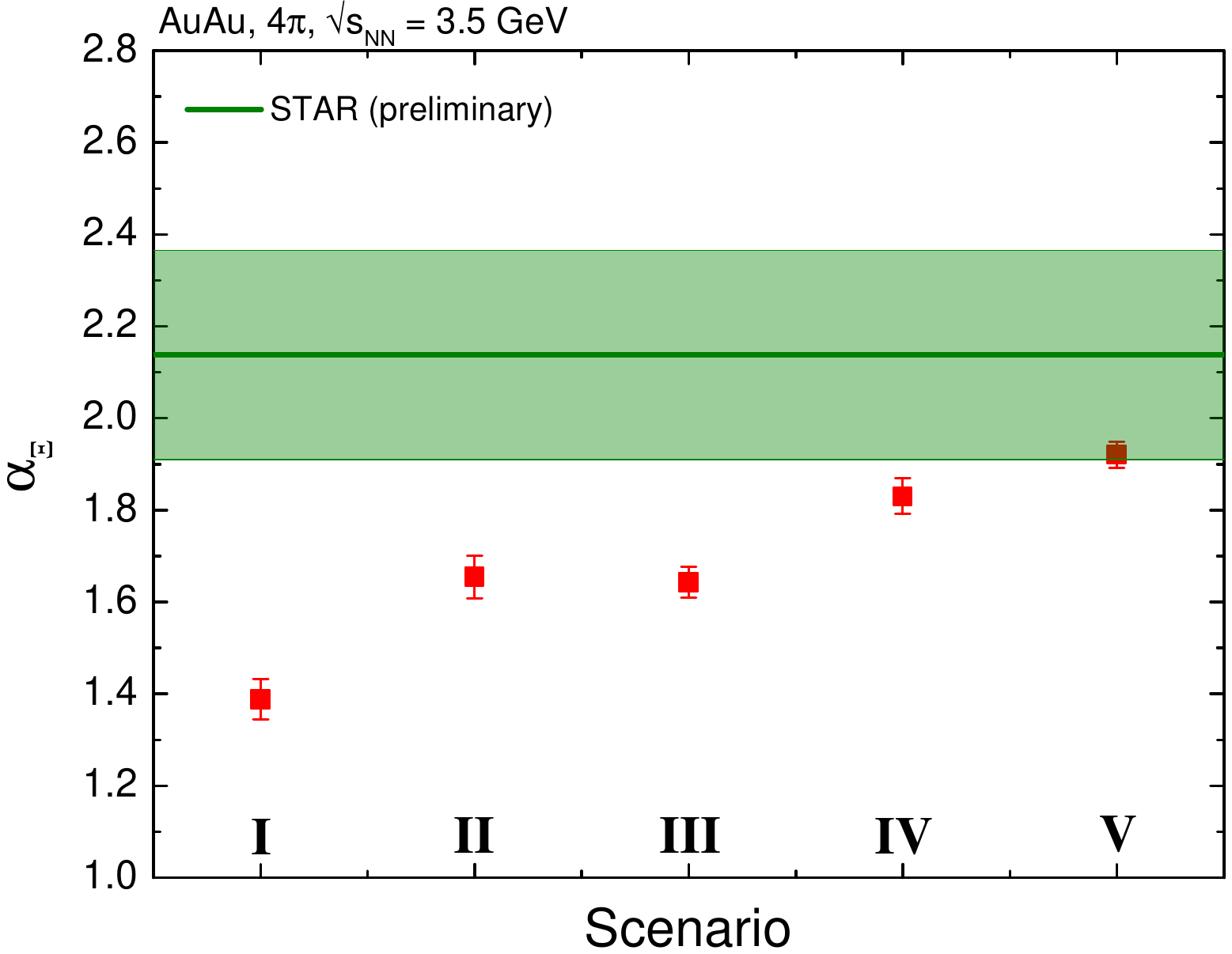}
    \caption{The $\alpha$ parameter for the $\Xi$-baryon at a beam energy of $\sqrt{s_{\mathrm{NN}}}=3.5$ GeV for five different scenarios. The scenarios are described in the text. The green shaded area corresponds to the preliminary STAR results at that energy.}
    \label{fig:scenarios}
\end{figure}

Even though we have clearly demonstrated that the $\alpha$ parameter can be a sensitive probe of the equation of state at different beam energies, it also became clear that the equation of state cannot be the only factor in determining $\alpha$. This is especially obvious in the case of the $\Xi$ baryon where neither equation of state, in connection with UrQMD v4.0 is able to describe the preliminary STAR data. It is therefore worthwhile to systematically investigate the other inputs of the model which may influence the parameter.
In the following, we will compare five different scenarios for which we extracted $\alpha$ for the $\Xi$ at a beam energy of $\sqrt{s_{\mathrm{NN}}}=3.5$ GeV. The results for these 5 scenarios are shown in figure \ref{fig:scenarios}.

\begin{enumerate}
\item Scenario I: This is the standard version of UrQMD v4.0 with the CMF equation of state. It includes production of the $\Xi-$baryon either via resonance excitation and decay or color string decay as discussed in section \ref{production}. We only allow for meson-baryon strangeness exchange reactions as shown in \cite{Graef:2014mra}.  
\item Scenario II: Here, a new scattering process, the baryon-baryon strangeness exchange is included. This process allows for the exchange of a strange quark between the $\Xi$ and nucleon, e.g. $\Xi + N \leftrightarrow Y +Y$. The cross section for this process is not well known and therefore this process is not switched on per default in UrQMD. In this scenario, we use the cross section as described in \cite{Graef:2014mra}. The effect of the process in this scenario will be either the reduction of the overall $\Xi$ yield in a case where the elementary production is large or the increase of the $\Xi$ multiplicity if the elementary production is small. Both effects may lead to an increase in $\alpha$.
\item Scenario III: In this scenario only the production cross section of the $\Xi$ via the excitation of a baryonic resonance is reduced and the string excitation is the dominant channel. No  $\Xi + N \leftrightarrow Y +Y$ exchange is allowed. Again, this leads to an increase in the $\alpha$ value as the other available meson-baryon strangeness exchange reactions now lead to an increase of the $\Xi$ in central collisions with respect to the overall reduced direct production cross section.
\item Scenario IV: Combining II and III, this scenario starts with a decreased overall cross section but allowing meson + baryon as well as baryon + baryon strangeness exchange reactions. This means that an even larger amount of $\Xi$ is produced in secondary exchange reactions and thus the sensitivity on the density of the created system increases.
\item Scenario V: This final case is similar to IV but is simulated in the cascade mode, i.e. with a much softer equation of state. The difference between IV and V shows again the small effect of the equation of state as compared to the other factors.
\end{enumerate}

The results for each of the above scenarios represents a specific choice for some poorly constrained input to the model. This shows that the actual uncertainties to the parameter are actually substantial \footnote{Note also a previous study which investigated the uncertainties in strangeness production fro unknown resonance properties \cite{Gerhard:2012fj}.}. 

Nevertheless, comparing these different scenarios leads to a clear conclusion. The $\alpha$ parameter, characterizing the centrality dependence, shows a strong influence on the relative effect of direct production of the strange baryons relative to possible secondary interactions which influence the relative yields of strange hadrons. It is therefore essential to have a precise input, e.g. for the fundamental production cross section in $p+p\rightarrow \Xi + X$ just above its production threshold, as we have for Kaon and $\Lambda$ production. Without this input precise conclusions can not be drawn and the interpretation of the data will remain ambiguous. Furthermore, specific strangeness exchange cross sections ($\Lambda \Lambda \rightarrow \Xi N$ or many-body interactions etc.) may be underestimated or even be strongly modified in the medium. More speculatively, at some point deconfinement may become important which is not treated in our current approach but has been discussed in the context of strangeness production in previous works \cite{Steinheimer:2011mp,Linnyk:2010cr}.

One should also note that one reason for the sensitivity to the different mechanisms is that STAR measures the $\Xi$ production above and only slightly below the threshold. Significantly below the $\Xi$ threshold some of these processes would be strongly suppressed and the enhancement due to the equation of state becomes more clear.

In the future, a similarly detailed comparison of the centrality dependence of charmed hadrons with SIS100 data may be possible. Unlike for the $\Xi$ baryon, for the charmed hadrons, to make a D-meson from a $\Lambda_c +\pi \rightarrow D + N$ exchange, a large amount of extra energy is required which will suppress this reaction. On the other hand, the creation of a $\Lambda_c$ from a D-meson (via $D +N \rightarrow \Lambda_c + \pi$) is strongly favored, but at beam energies below the elementary threshold, the production probability to produce a D-meson is also suppressed by two orders of magnitude compared to the $\Lambda_c$, which makes these kind of reactions unlikely.

\section{Summary}

We have discussed the opportunities for constraining the high density QCD equation of state with charm measurements at the upcoming SIS100 accelerator. Due to the high event rates expected at the CBM experiment, the new facility will allow for the first time the measurement of multi-strange hadrons as well as charmed hadrons in heavy ion reactions at energies below the elementary threshold energies. Similar to results at lower beam energies, where the production of Kaons was sensitive to the equation of state at low energies, the production cross section of $\Xi$, $\Omega$ and charmed hadrons are sensitive to the equation of state reached at higher beam energies. 

On the other hand, we have also discussed the importance of having a solid baseline for the production cross section of these strange and charmed hadrons in elementary p+p collisions. Without these cross sections it will be almost impossible to disentangle the different aspects of strangeness and charm production in heavy-ion reactions. Fortunately, the FAIR facility will also provide opportunities for such baseline studies. 

A dedicated baseline measurement campaign to study strangeness and charm production at the SIS100 in p+p and p+A collision systems to pin down the elementary production cross sections, is a prerequisite for a conclusive interpretation of the promising but complex A+A data. 
These A+A collisions would then provide very important constraints for the QCD equation of state of the system created in these collisions.

\section*{Acknowledgments}
The computational resources for this project were provided by the Center for Scientific Computing of the GU Frankfurt and the Goethe--HLR and GSI green cube. The publication is funded by the Open Access Publishing Fund of GSI Helmholtzzentrum fuer Schwerionenforschung.\\

This work is theoretical research and the data sets generated during the current study are available from the corresponding author on reasonable request.

\bibliography{sn-bibliography}

\end{document}